# Liquid Atomization out of a Full Cone Pressure Swirl Nozzle

Nicolas Rimbert and Guillaume Castanet

[*] Nancy University, LEMTA,

2 av. de la Forêt de Haye, F-54504 Vandoeuvre-lès-Nancy cedex

nicolas.rimbert@esstin.uhp-nancy.fr and guillaume.castanet@ensem.inpl-nancy.fr

**Keywords**: Liquid atomization, Spinning jets, Instability theory, Turbulent atomization, Lévy stable statistics

**Abstract**

A thorough numerical, theoretical and experimental investigation of the liquid atomization in a full cone pressure swirl nozzle is presented. The first part is devoted to the study of the inner flow. CAD and CFD software are used in order to determine the most important parameters of the flow at the exit of nozzle. An important conclusion is the existence of two flow regions: one in relatively slow motion (the boundary layer) and a second nearly in solid rotation at a very high angular rate (about 100 000 rad/s) with a thickness of about $4/5^{th}$ of the nozzle section. Then, a theoretical and experimental analysis of the flow outside the nozzle is carried out. In the theoretical section, the size of the biggest drops is successfully compared to results stemming from linear instability theory. However, it is also shown that this theory cannot explain the occurrence of small drops observed in the stability domain whose size are close to the Kolmogorov and Taylor turbulent length scale. A Phase Doppler Particle Analyser (PDPA) is used to characterize the droplet size and velocity distribution. Due to centrifugal force, the smaller droplets tend to prevail on the spray axis and a peak close to the Taylor length scale appears progressively in the PDF when increasing the distance from the nozzle. It is then assumed that these small droplets are the results of a turbulent cascading atomization process and that in the near nozzle area, since centrifugal segregation is small, PDF can therefore be fitted with a log-stable law (Rimbert and Séro-Guillaume, 2004). The value of the stability index is found to be 1.35 very close to a previous experimental result of 1.39 (Rimbert and Delconte, 2007) but far from a known theoretical value of 1.70 (Rimbert, 2010). This let think about a slightly different underlying process due to the helical nature of the turbulence.

## 1- Introduction

Instability theory is generally used to describe the large scale behavior of the spray. The most important mechanism in high rotating speed, full cone, pressure swirl atomizer is the centrifugal force which can be crudely associated to the classical Rayleigh-Taylor instability [1]. Recently, Kubitschek and Weidman [2,3] made an important contribution by making the full linear stability analysis of a helical column of fluid in zero gravity and by successfully comparing their prediction to experimental data collected through high-speed imaging. However, the relevance of this kind of approach can be questioned far away from the instability thresholds. For instance, in the present study, the liquid within the swirl nozzle has an angular velocity about three orders of magnitude higher than in the previously mentioned works of Kubitschek and Weidman. This observation leads the way to a second possible mechanism of atomization as a process driven by turbulence.

Modern reviews on atomization can be roughly divided into two categories: turbulent atomization based on Kolmogorov work [4] vs. linear instability/ligament mediated atomization [5]. The present work aims at proving that large droplets are mainly governed by classical linear instability theory while small droplets are mainly produced by the turbulence developing in the mixing zone. The shape of the droplet size distribution is shown to be a logarithmically stable law (cf. [6]) which can be related to turbulent intermittency modeling (cf. [7,8,9,10,11]). While akin to ligament mediated atomization [12,13], the resulting scenario makes use of classical turbulent quantities (turbulent kinetic energy density $k$ and dissipation $\varepsilon$) rather than to new empirical parameters.

This article begins with a description of the experimental facility as well as the geometry of the simplex pressure swirl nozzle under investigation. A correct order of magnitude for the angular velocity of the spray can be obtained from the conservation of the angular momentum and the result compares well with Computational Fluid Dynamics. These calculations can then be used as starting point for Kubitschek and Weidman linear stability analysis and it will be shown that only the size of the largest droplets can be related to this mechanism: droplets smaller than 50 $\mu m$ cannot be produced by this mechanism and there are experimental evidences that they are numerous. Therefore their size distribution is studied in the last section of this paper in relation to turbulent intermittency modeling.

## 2- Description of the Inner flow

The nozzle considered in this study is a simplex pressure swirl nozzle from Danfoss (OD 1.87Kg/h EN 60°I 500 0.50 US gal/h 45°S) that is fed by pressurized water at 6 bars. Figure 1 presents the inner geometry of the nozzle: a funnel, located just before the nozzle exit, is connected to three square channels giving to the flow an angular momentum respective to the $z$ axis. The 10:1 contraction ratio of the funnel thereafter enhances the rotation speed. This is a full cone nozzle since the outflow radius is small enough not to allow some outside air to penetrate through the $z$ axis inside the nozzle, in contrast with hollow cone pressure swirl nozzle.

*(a)*            *(b)*

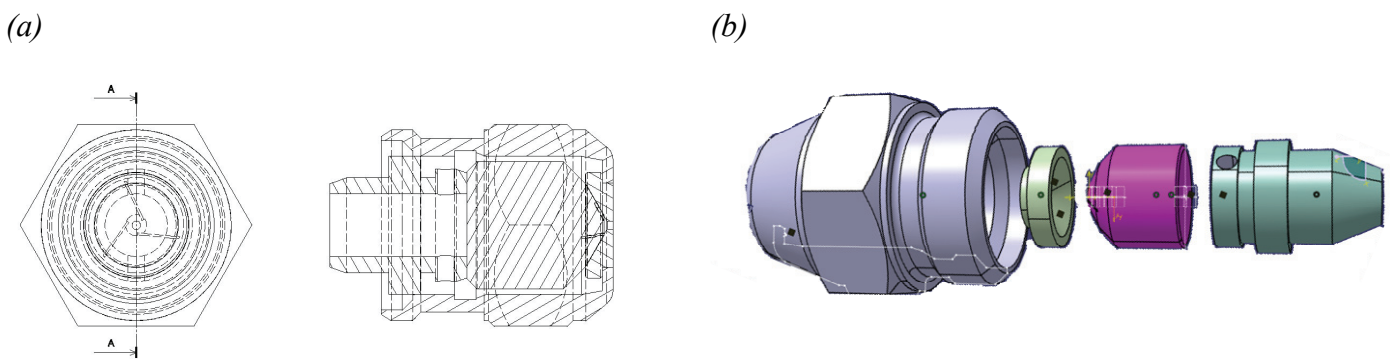

**Figure 1: sketch of a simplex pressure swirl nozzle. (*a*) the funnel is fuelled by three channels forming a 120° angle. These channels are off axis conferring a *z*-angular momentum to the liquid. This rotation is amplified by the funnel. Arrows shows the path of the liquid, (*b*) CAD representation of the four parts of the nozzle).**

In a first simplified approach, the conservation of the angular momentum can be used to estimate the angular velocity $\omega$ of the liquid at the exit of such an injector. By defining the axial angular momentum $J_z$ around the vertical $z$ axis (cf. Figure 2) of liquid inside the control volume $V_c$ as:

$$J_z = \mathbf{J}.\mathbf{e_z} = \int_{V_c} (\mathbf{OM} \wedge \rho_L \mathbf{V}).\mathbf{e_z} \, dv, \tag{1}$$

where $\rho_L$ is the liquid density, $\mathbf{V}$ the liquid velocity and $\mathbf{OM}$ a position vector (cf. Figure 2). By balancing inflow (thereafter subscripted $i$) and outflow (thereafter subscripted $o$) vertical angular momentum, the following equation can be written:

$$\frac{dJ_z}{dt} = \int_{V_c} \frac{\partial(\mathbf{OM} \wedge \rho_L \mathbf{V}).\mathbf{e_z}}{\partial t} dv + \underbrace{\iint_{\Sigma_e} (\mathbf{OM} \wedge \rho_L \mathbf{V}).\mathbf{e_z}(\mathbf{V}.\mathbf{n}) ds}_{\dot{J}_{z,i}} - \underbrace{\iint_{\Sigma_s} (\mathbf{OM} \wedge \rho_L \mathbf{V}).\mathbf{e_z}(\mathbf{V}.\mathbf{n}) ds}_{\dot{J}_{z,o}}. \tag{2}$$

Assuming a steady flow, the later expression can be simplified leading to:

$$\dot{J}_{z,i} = \dot{J}_{z,o}. \tag{3}$$

In the inflow, a vertical angular momentum is released to the flow because there is a certain distance between the vertical axis and the midline issued from the three channels. This distance is called the eccentric length $e$ while the channel square section is denoted $A$. From the expression of $\dot{J}_{z,i}$ in eq.2,

$$\dot{J}_{z,i} = 3e\rho_L A V_i^2. \tag{4}$$

At the nozzle exit, the outflow can be seen as a cylindrical jet having a radius equal to $r_o$ and an axial and angular velocity respectively $V_{z,o}$ and $\omega$: the radius of the nozzle outlet. Using eq.2, it comes that

$$\dot{J}_{z,o} = \frac{1}{2}\pi\rho_L r_o^4 \omega V_{z,o}, \tag{5}$$

Finally, using eq.3 and the parameters given in table 1, the following value is obtained for the angular velocity $\omega$:

$$\omega = \frac{6eAV_i \cos(\theta)}{\pi r_o^4} \approx 410,000 \text{ rad.s}^{-1} \tag{6}$$

**Table 1: geometric parameters of the nozzle**

| Parameters | Value |
|---|---|
| Outlet radius $r_o$ | 106 $\mu$m |
| Inlet channel side length ($A^{1/2}$) | 108 $\mu$m |
| Inlet velocity $V_i$ | 18 m/s |
| Inlet angle $\theta$ | 60° |
| Eccentric length $e$ | 260 $\mu$m |

As friction losses are neglected in this calculation, the outflow appears to be rotating at a tremendous speed. At such speed boundary layers in the nozzle cannot be neglected anymore and a more refined calculation is then required based on a full simulation of the inner flow that resorts to Computational Fluid Dynamic. This was performed using Fluent as described more thoroughly in [1]. Qualitative results of these computations, as well a sketch of the computational domain, can be found in Figure 2 where some sample pathlines are depicted. More quantitative results are summarized in table 2.

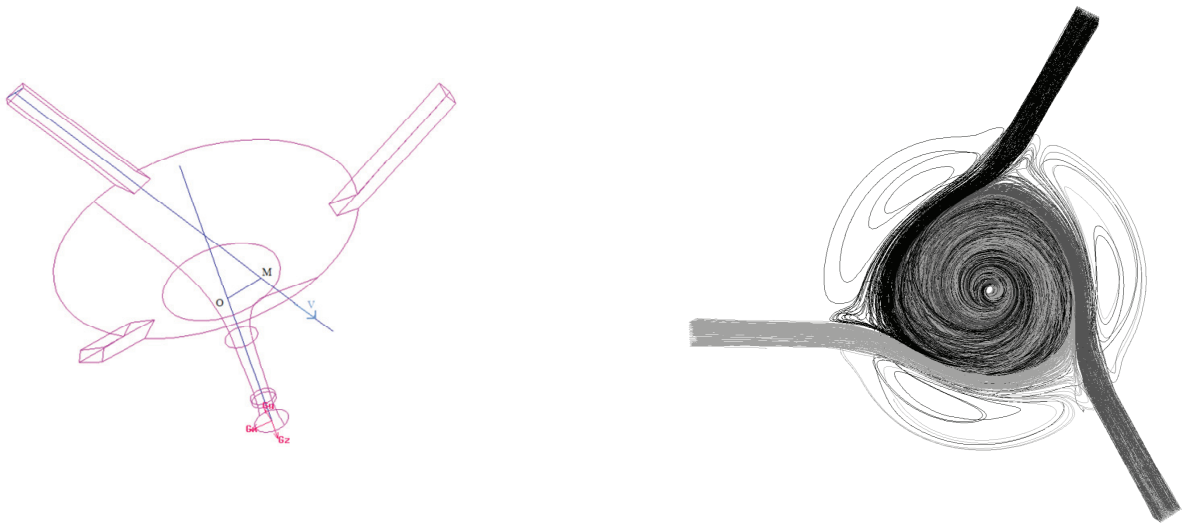

**Figure 2: sketch of the computational domain used in the CFD analysis (a) and sample pathlines (b) obtained from the numerical simulation of the nozzle inflow (color is a tag of the different liquid channels)**

**Table 2: Average outlet parameters of the Danfoss swirl nozzle, inlet pressure is 5 bars**

| Parameters | Value |
|---|---|
| Axial velocity $Vz$ | 18 m/s |
| Angular velocity $\omega$ | 100,000 rad.s$^{-1}$ |
| Turbulent kinetic energy $k$ | 14 m$^2$/s$^2$ |
| Turbulent dissipation $\varepsilon$ | 10$^6$ m$^2$/s$^3$ |

It can be seen in Figure 3 and in table 2 that the mean angular velocity obtained in numerical simulation is much less than the velocity predicted in eq.6. This means that inner frictions are very important inside the nozzle resulting in a torque which may act on the screw that fastens the nozzle elements (Figure 1). Even when frictions are taken into account, the angular velocity remains very high corresponding to a radial acceleration $a$ in the order of 1,000,000 m/s$^2$, i.e. 100,000g. Such a high angular velocity is only reachable for very small apertures (here around 200 $\mu m$) without leading to unrealistic velocity (as far as the flow is considered uncompressible).

This important radial acceleration is the main atomizing mechanism in this kind of nozzle. However, the appearance of a large boundary layer is an inhibiting factor for atomization since it slows down nearly 25% of the liquid.

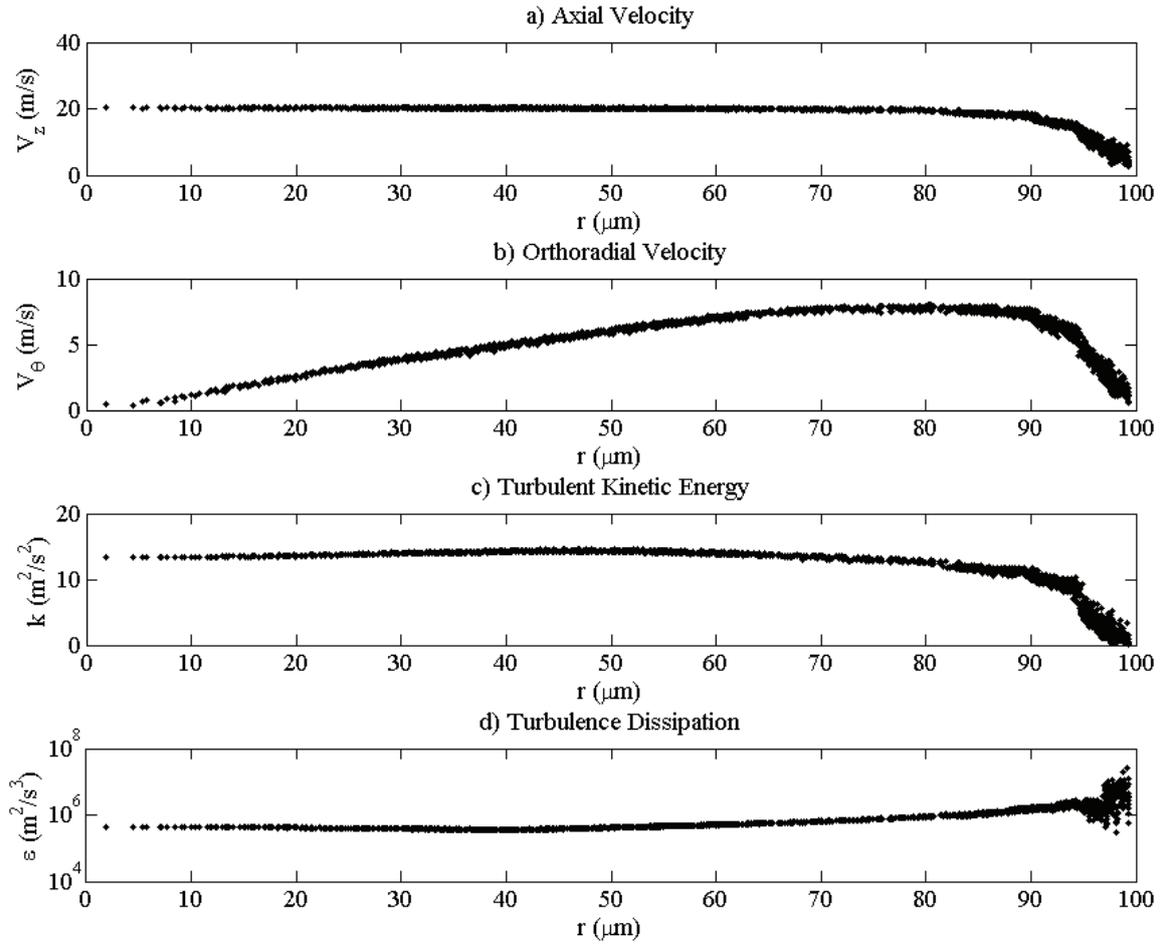

**Figure 3: Axial, orthoradial velocity and turbulent quantities at the nozzle output versus radial position. Fluent's RNG formulation adapted to highly swirling flow [14] has been used but it also compares well with classical RSM model such as Launder, Reece and Rodi [15].**

### 3- Description of the outer flow

A high-speed camera (FASTCAM-ultima APX RS Photron ) was used to observe the spray. Shadow images were taken at 75000 frames per second. In figure 3, a cone can be clearly seen below the nozzle exit. This cone is in fact the extension of the boundary layer that develops within the nozzle body (section 2). The cone is not hollow as it contains a mist of droplets resulting from the atomization of the main vortex core. As presented in Figure 4, the outer cone breakup length $L$ increases significantly with the injection pressure. From these images, this length was estimated to 1.8 *mm* for an injection pressure of 6 bars. This value is close to the prediction by Dombrowski and Hooper's empirical law [16] which gives

$$L = 12 \frac{U}{\omega} = 12 \frac{17}{100,000} \approx 2 mm .\qquad(7)$$

However, this value may be underestimated because the outer sheet is rotating slower than the inner core, thanks to inner friction inside the nozzle. Moreover, high-speed imaging has shown the sheet breakup length to be a very fluctuating quantity therefore some caution has been taken when defining the first measurement point.

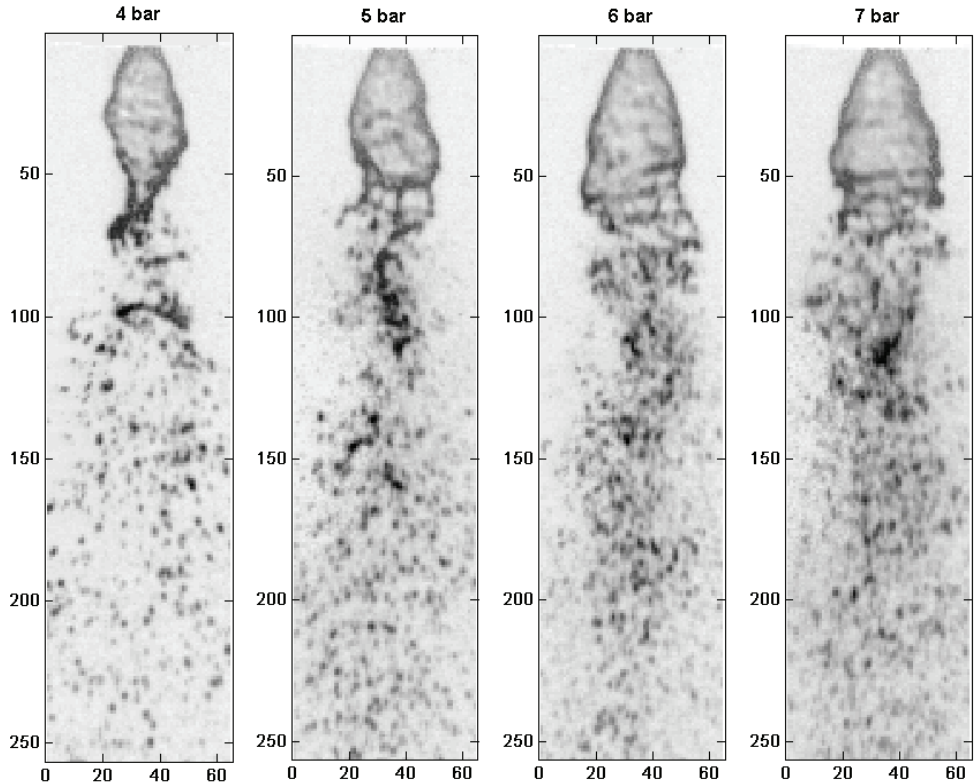

**Figure 4 Images from the high-speed imaging using a FASTCAM-ultima APX RS Photron camera at 75000 frames per second (scale is given in pixels 1 pixel ≈ 30 $\mu m$)**

In addition to high-speed imaging, a Phase Doppler Particle Analyzer (PDPA) from Dantec Dynamics was used to investigate the spray. Droplets are illuminated by a system of two crossing laser beams obtained from the green line (wavelength: 514.5 nm) of an $Ar^+$ laser. The measurement volume formed by the intersection of the laser beams is roughly an ellipsoid of about 146 x 146 x 2003 $\mu m^3$. From the PDPA measurements, it is possible to determine the size and velocity PDF of the droplets passing through the measurement volume and also to establish joint size-velocity distributions. Only the vertical component of the droplet velocity, parallel to the direction of injection, is measured with the present system. The PDPA receiver is positioned in the forward scattering direction under a scattering angle of 50° so that the $1^{st}$ order refraction mode is dominant. The maximum droplet size that can be measured in this configuration is 210 μm. This limit results from considerations on the phase of the detector signals to avoid a 2π-ambiguity [17]. However the intensity of the light scattered by a droplet is roughly proportional to its surface. Therefore, at fixed threshold levels and high voltages applied to the photomultiplier tubes (PMT), bigger droplets may saturate the PMT while the smaller ones may not trigger the detection. In practice, this behavior reduces the effective range of the droplet size measurement. According to the PDPA manufacturer, droplets can be detected and processed if their diameter ranges from 1x to 40x (or 1.6 decades). In the present case, this limitation does not appear to be a major drawback. The detection and validation rates are sensitive to many parameters including the spray density, the droplet sphericity and trajectory in the measurement volume [17], therefore measurements of mass fluxes are rather inaccurate with this technique. For this reason, only PDF are provided in the present analysis of the spray.

To avoid spurious reflection of the laser on the outer sheet as well as non-spherical droplets, measurements have been made starting 6 *mm* downward (this is three times the average value predicted by eq. 19 but the angular velocity of the rotating sheet appears to be a very fluctuating quantity and can slow down greatly which according to eq. 19 greatly increase the breakup length). The PDPA moved on a grid and results are depicted in Figure 5 for the mean droplet size $d_{10}$ and the mean droplet velocity. Near the exit, the average droplet velocity can be estimated to be 17 m/s close to the CFD estimate of 18 m/s. There is a clear segregation: large droplets are mainly on the border of the cone as they are more influenced by the centrifugal acceleration than by the air friction.

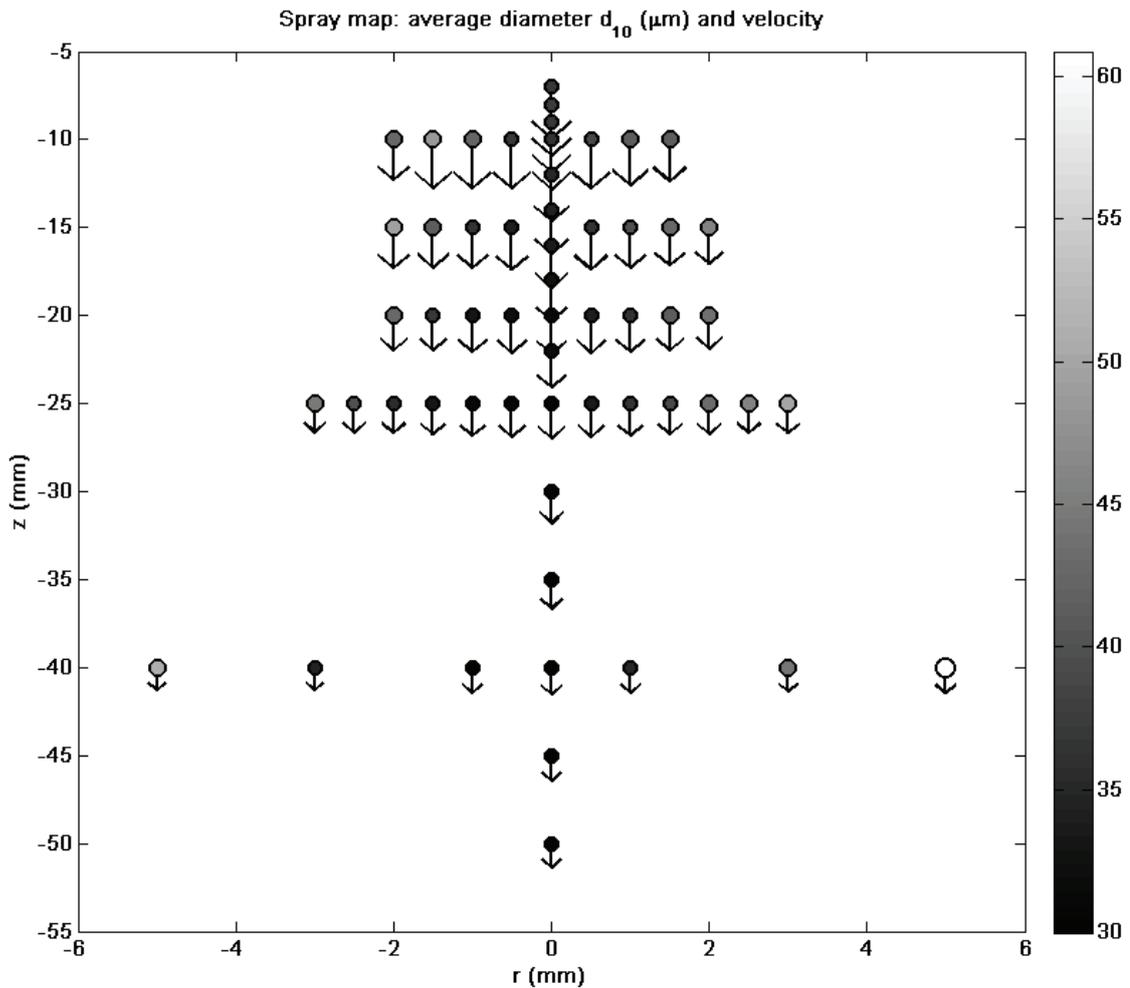

**Figure 5 Spatial extension of the spray. Centrifugal forces tend to select the biggest droplets on the border of the spray. Axial scales are given in *mm* and mean droplet diameter $d_{10}$ range from 30 to slightly over 60 *μm*; arrow length are proportional to the mean droplet velocity. Maximum mean velocity is found to be 17 m/s.**

Despite all the droplets undergo a centrifugal acceleration, the effect of segregation is related to the fact that the deceleration due to the drag force is larger for smaller droplets leading to a higher proportion of larger droplets on the edge of the spray.

This effect can be seen more precisely on the joint size-velocity PDF depicted in Figure 6. Small droplets, in the order of 10 microns, have an average velocity close to 10 m/s whereas larger droplets are flowing with almost the inlet speed i.e. 17 m/s. Note that the value scale is in logarithmic and that the measurement point is located 30 mm under the nozzle exit.

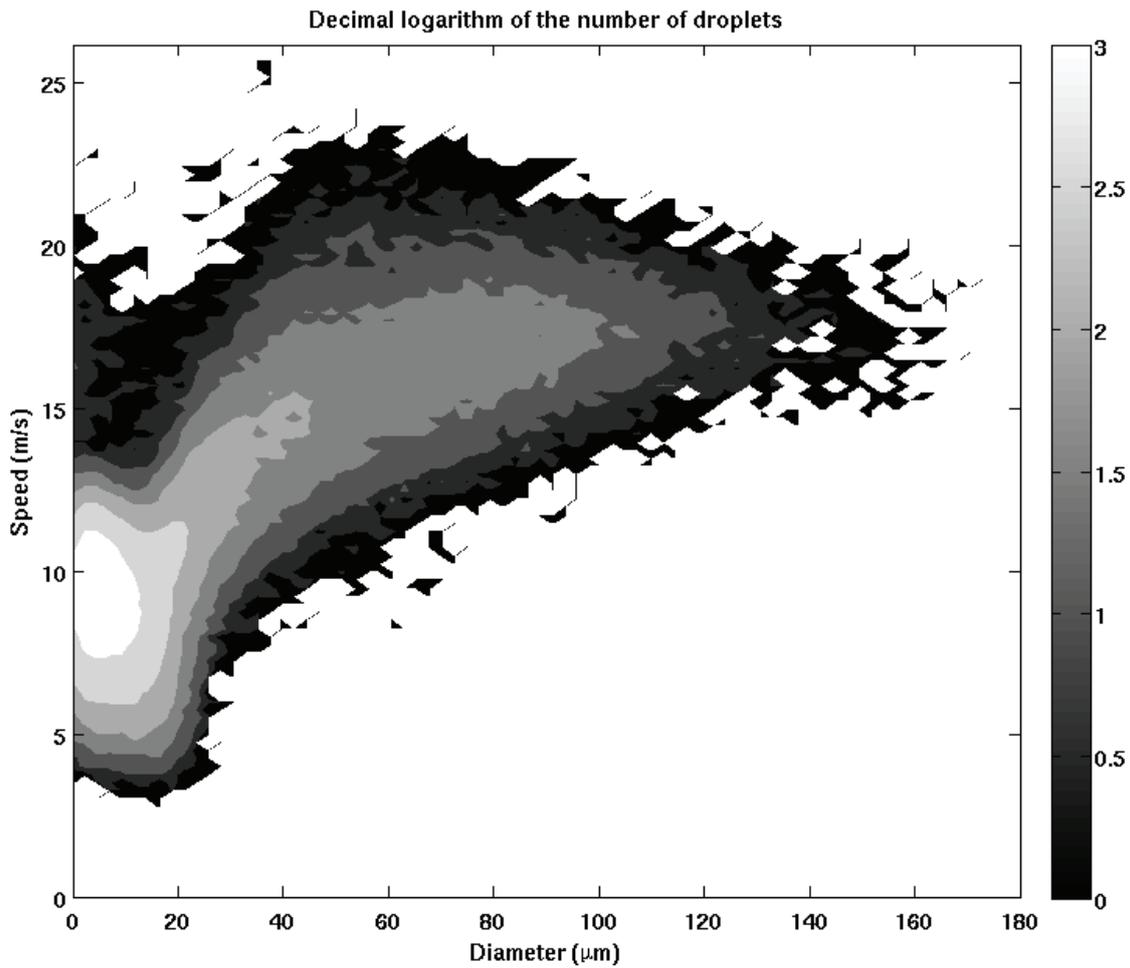

**Figure 6: Joint size-velocity PDF of the spray 30 mm downward. Injection gauge pressure is equal to 6 bars. Values are related to the number of detected droplets during the acquisition of a sample of 50 000 droplets (in decimal logarithmic scale)**

To make a proper assessment of the droplet PDF stemming from the primary atomization of the spray, it is necessary to correct the data from the segregation effect occurring upstream in the spray. However, this correction is almost unnecessary if the data are collected very close to the nozzle exit. Figure 7 presents the joint size-velocity PDF obtained for the measurement point that is the closest from the injector, i.e. 6 mm downstream compared to the measurements made previously at 30 mm. To obtain this figure, the joint size-velocity PDF has been sampled with 80 different size bin. For each size bin, a Gaussian PDF of the droplet velocities has been fitted from the conditional velocity PDF. The resulting average velocities are plotted versus the droplet diameter in Figure 7 with lines (continuous: 6mm, dashed: 30 mm) while the standard deviation is represented by symbols (dots: 6mm, triangles: 30mm) The only difference between these two two curves is that at 6mm, two millions droplets were processed to insure a better statistical convergence when calculating the averaged velocity corresponding to each diameter class and the size marginal distribution. From the comparison between data measured at 6 mm and 30 mm, it can be seen that the slowdown of small droplets is much lessened at 6mm down the exit although it cannot be fully neglected.

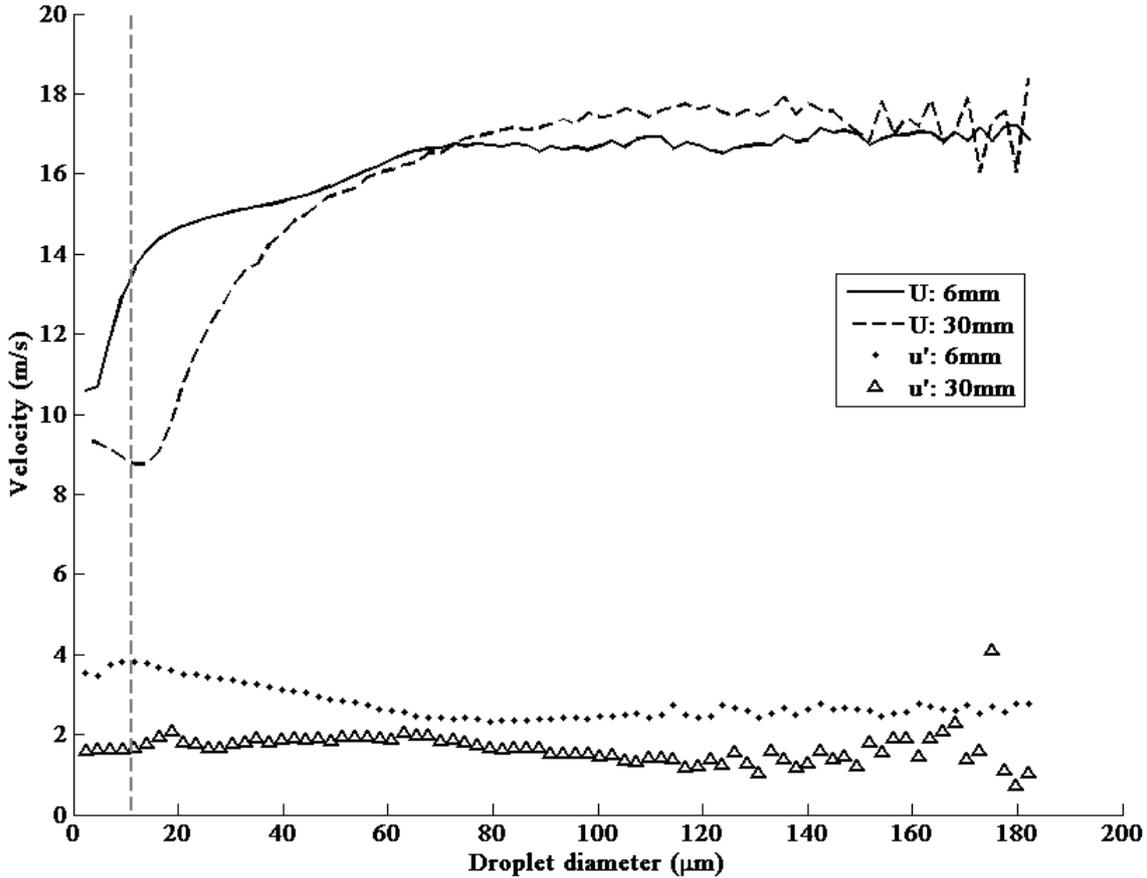

**Figure 7: Estimated centerline of the joint size-velocity PDF, black triangle are experimental mean data, gray dots stand for the standard deviation around this value and the dashed line is the theoretical limit (12) for droplet drag. Data are collected 6mm downward from 2,000,000 droplets. Injection gauge pressure equals 6 bars.**

To figure out how much the segregation effect can modify the droplet size-velocity distribution on the spray vertical axis, a simple analytical model has been developed. It arises from the momentum conservation applied to a droplet:

$$\begin{cases} \dfrac{dv_z}{dt} = -\dfrac{3}{4} \dfrac{\rho_G}{\rho_L} \dfrac{1}{d} C_D (\mathrm{Re}) v^2 \\ v_z(0,0) = v_z^0 \approx 17 \text{ m.s}^{-1} \end{cases} \quad (8)$$

In this equation, only the vertical component of the droplet velocity is considered because only droplets on the spray axis are of interest. Near the exit of the injector, the acceleration of gravity plays a limited role due to the droplet inertia. This system can be easily integrated for low Reynolds numbers, as the drag coefficient can be estimated from the Stokes' drag law, $C_D = 24/Re$. This leads to:

$$v_{z,1} = v_{z,0} \exp\left(-\frac{t_1 - t_0}{\tau_{drop}}\right), \qquad (9)$$

where $\tau_{drop}$ is a characteristic relaxation time for the droplets to reach their terminal velocity.

$$\tau_{drop} = \frac{d^2}{18\nu_G}\frac{\rho_L}{\rho_G}. \qquad (10)$$

If this relaxation time is large compared to the transit time between the exit of the nozzle and the position of the probe, then the droplet velocity can be considered to be unaffected by air friction. Let us compute this limit:

$$\tau_{drop} \gg \frac{z_{probe}}{V_z} \qquad (11)$$

$$d \gg \sqrt{\frac{\rho_G 18 \nu_G z_{probe}}{\rho_L V_z}} \approx \sqrt{\frac{1.3 \times 18 \times 1.5 10^{-5} \times 6 10^{-3}}{1000 \times 18}} \approx 11 \mu m \qquad (12)$$

It appears that droplets having a diameter less than 11 μm may be stopped by air friction before reaching 6 mm downstream, i.e. the position of the first measurement. This limit is superimposed in Figure 7 as a dashed line.

## 4. Modeling of the atomization

As already stated, models on atomization can be divided into two categories: turbulent atomization and instability mediated atomization. In this section, these two approaches are used in order to assess the instability wavelength, the turbulent length scales that are normally in close relation with the size of the released droplets. Later, in section 5, the results of these analyses will be compared to the experimental size distributions obtained with the PDPA. As a preliminary remark, it should be pointed out that the breakup mechanism cannot be presently the air friction, since the Weber number and the Ohnesorge number have rather low values: $We = \frac{\rho_G U^2 d}{\gamma} \approx 1$ and $Oh = \frac{\mu_L}{\sqrt{\gamma \rho_L d}} \approx 8.10^{-3}$ [18, 19].

### 4.1. Turbulent length scales

Given the high rotation velocity in the nozzle (about 100,000 rad.s$^{-1}$), turbulence cannot be ruled out as a secondary mechanism for the atomization. Using the results of the internal flow simulation, presented in section 2, it is possible [1,20] to obtain an order of magnitude for the main turbulent length scales. This includes:

- The integral scale:

$$L_{int} \cong \frac{V u'^2}{\varepsilon} \cong 168 \mu m \quad \text{(magnitude 2.23)} \qquad (13)$$

As expected, it can be noticed from this value, that the boundary layer is surrounding the main vortex core in the interior of the nozzle, i.e. $d_{nozzle} \approx \theta + L_{int} + \theta$ with $\theta$ the boundary layer thickness close to 25 µm.

- The Taylor scale (average size of the dissipating eddies):

$$\lambda_T \cong \sqrt{15 \nu_L \frac{k}{\varepsilon}} \cong 14 \mu m \quad \text{(magnitude 1.15)} \qquad (14)$$

- The Kolmogorov length scale:

$$\eta \cong \left( \frac{\nu_L^3}{\varepsilon} \right)^{1/4} \cong 1 \mu m \quad \text{(magnitude 0.00)} \qquad (15)$$

All these length scales are also given in decimal magnitude with magnitude 0 standing for the Kolmogorov scale, or here, 1 micrometer. These values will be particularly useful for the forthcoming comparison with the experimental results.

### 4.2. Instability Theory

In [1] assumption was made that a slight modification of the classical Rayleigh-Taylor analysis would allow determining the maximum amplified wavelength in presence of an additional centrifugal force. In this analysis, since the centrifugal acceleration $a$ supersedes the terrestrial gravitational acceleration *g* classically used, this latter was simply replaced by the centrifugal acceleration. As a result, the wavelength of the Rayleigh-Taylor instability $\lambda_{RT}$ can be expressed as:

$$\lambda_{RT} = 2\pi \left( \frac{3\gamma}{\rho_L a} \right)^{1/2} \simeq 93 \mu m \quad \text{(magnitude 1.97)} \tag{16}$$

Although this value of $\lambda_{RT}$ may be a good estimate for the size of the larger droplets (section 5), it is based on the assumption of a plane liquid-gas interface that is inherent to the classical description of Rayleigh-Taylor instabilities. However, this assumption is irrelevant in the present case, since the liquid jet leaving the injector has initially an almost cylindrical shape. Moreover, as the radius of curvature of the liquid interface (roughly 106 $\mu m$, the diameter of the outlet orifice) is of the same order of magnitude as the wavelength, the influence of the curvature cannot be neglected. Finally the centrifugal acceleration, unlike the gravity, cannot be kept constant like in this crude description. It is actually increasing as the square of the distance to the rotation axe.

Recently, these limitations were overcome by Kubitschek and Weidman [2]. They made a very comprehensive study of the instabilities in the case of uniformly rotating liquid jets. The following analysis is directly based on their work. Two dimensionless numbers are of prime importance to describe the instability, the rotational Reynolds number

$$\text{Re}_{rot} = \frac{r^2 \omega}{\nu_L}, \tag{17}$$

and the so-called Hocking parameter [21]

$$L = \frac{\gamma}{\rho_L r^3 \omega^2} \tag{18}$$

which is actually the inverse of the Bond number. In our case $\text{Re}_{rot} \approx 1123$ while $L \approx 6.10^{-3}$. These values are close to some results given in [2] for $\text{Re}_{rot} = 1000$ and $L \approx 10^{-2}$ but making the full calculation gives some difference as can be seen on Fig. 8. In theses computation only the axisymmetric $n = 0$ mode is taken into consideration as it has been proven to be the most unstable mode.

Using these new results, the maximum amplified non dimensional wavenumber is found to be:

$$k_{max} = 7.2 \tag{19}$$

which leads to maximum amplified wavelength of

$$\lambda_{max} = \frac{2\pi r}{k_{max}} \approx 87 \, \mu m \quad \text{(magnitude 1.94)} \tag{20}$$

Interestingly enough, this is still of the same order of magnitude as the Rayleigh Taylor crude estimate. But more interestingly, the marginally stable wavenumber can be estimated, and one gets:

$$k_{marg} = 13 \tag{21}$$

so that the marginally stable wavelength is

$$\lambda_{marg} = \frac{2\pi r}{k_{marg}} \approx 48 \, \mu m \quad \text{(magnitude 1.68)} \tag{22}$$

meaning that shorter wavelength cannot be amplified. Therefore the appearance of droplets of size close to the micrometer cannot be explained by this mechanism (Note that neither droplets of the scale of the Kolmogorov length nor of the Taylor microscale can be predicted by the linear instability theory). Also note that these wavelength are not equivalent to the resulting droplet diameter and equating the volume of the destabilized liquid cylinder to the volume of the resulting droplet, $\pi/6 \, d^3 = \pi r^2 \lambda$, one easily gets the following relation

$$d = \left(6 r^2 \lambda\right)^{1/3} \tag{23}$$

which introduces a shift in the magnitude i.e. one gets 180 $\mu m$ (magnitude 2.26) for the diameter of the fastest growing droplet and 148 $\mu m$ (magnitude 2.17) for the marginally stable one (here $r$ =106 $\mu m$, assuming $r = L_{int}/2 = 84$ $\mu m$, leads to the respective value of 154 and 127 $\mu m$ or magnitude 2.19 and 2.10).

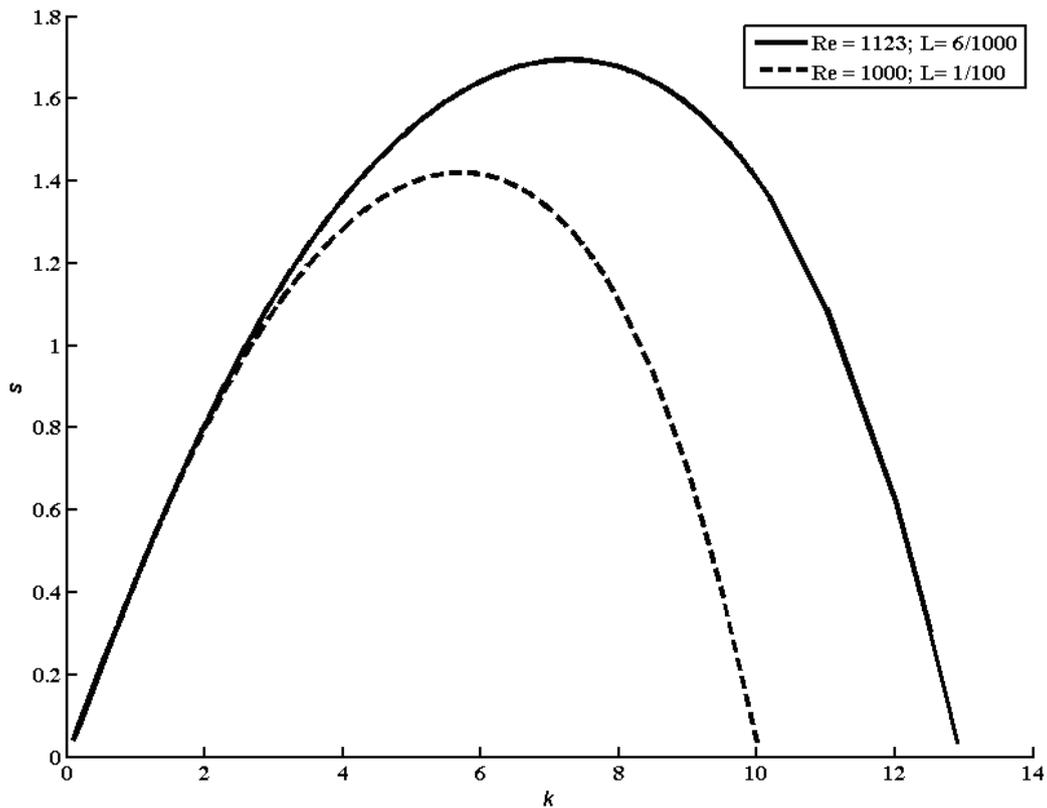

**Figure 8: Computation of the most amplified axisymmetric *n* = 0 wavelength in our experimental configuration using Kubitschek and Weidman linear analysis [2]. Comparison is made with the closest result given in [2] for Re = 1000 and L = 1/100. X axis is the non dimensional wavenumber k. Y axis the amplification rate.**

## 5. Results and discussion

Figure 9 represents the size marginal PDF at three different locations on the spray vertical axis (7, 16 mm and 30 mm) in a semi logarithmic scale. Reference values of the different characteristic scales of turbulence and instability theory are recalled for comparison. It can be seen that most droplets are actually located between the value of the Kolmogorov scale (1 μm, magnitude 0) and the nozzle diameter (210 *μm*, magnitude 2.3). Two peaks are present, the first one being located close to the Taylor micro-scale (Eq.14) and the second one, much less visible, close to the droplet diameter predicted by the instability theory (Eq.23). The peak corresponding to the Taylor micro scale becomes more prominent with the distance to the nozzle exit.

This does not mean that small droplets (in the order of the Taylor scale) are progressively created from the breaking of larger droplets since there is no secondary atomization (see the remark at the beginning of section 4). Taylor sized droplets being the more numerous; this is another illustration of the aforementioned segregation effect, larger droplets being ejected outward.

From figure 9, it appears that the prediction of the model based on the instability theory is not conclusive. At best, it can predict the sizes of the bigger droplets but it cannot explain the appearance of most droplets that are smaller than about one hundred micrometers.

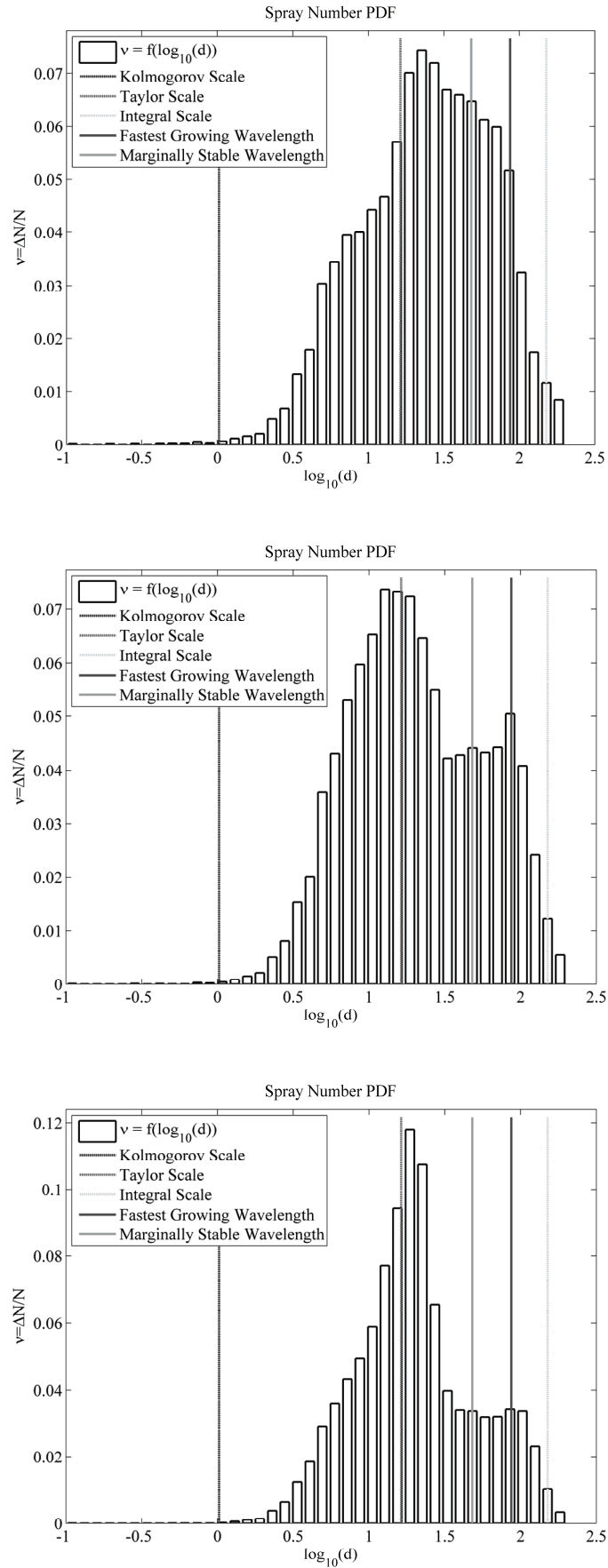

**Figure 9: droplet size PDF at 7 mm, 16 mm and 30 mm down to the spray nozzle. Data are collected on 50,000 droplets. The spray magnitude PDF is well located inside the interval defined by the Kolmogorov scale and the integral scale. Most common droplets are very close to the Taylor scale**

Since another mechanism must be at work, the assumption of a fragmentation assisted by turbulence will be studied more in detail in the following. To that end, the analysis of the size PDF at 6 mm is limited to an interval corresponding to the magnitudes [0.8, 1.9]. In this manner, on one hand, small droplets whose occurrence has been strongly modified due to the segregation effect are not taken into account. A cutoff at 0.8, i.e. about 6 microns, was chosen. It corresponds to the size of the droplets undergoing a reduction by 50% of their velocity according to the model established in chapter 3. One the other hand, droplets of size greater than magnitude 1.9 are excluded because their formation can be explained by other mechanism than turbulence intermittencies, namely instabilities.

The figure 11 presents the result of the fitting of the size PDF at 6mm with a log-stable law. There is an almost perfect matching between the fitted curve and the experimental data. Log-stable laws are considered here because they are widely used in the modeling of turbulent intermittencies. Log-stable laws are known as possible asymptotic solutions to self similar fragmentation equations [6]. However, no clear relationship between atomization and turbulent intermittencies has been drawn yet in the literature.

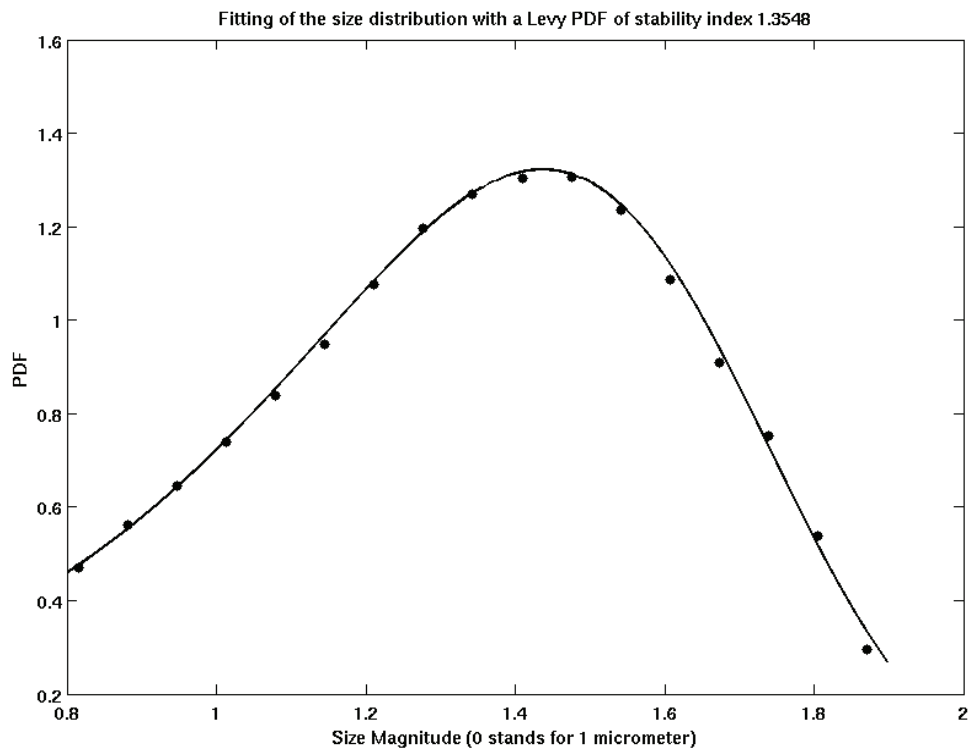

**Figure 10: Marginal size PDF of the spray 6mm downward. Fit of a log stable law. Resulting parameters are $\alpha$ = 1.35, $\beta$ = -1, $\sigma_{\log d}$= 0.38 ($\sigma_{\ln d}$= 0.88), $\delta_{\log d}$ = 0.93. Data are collected on 2,000,000 droplets.**

In [11], the log-stable intermittency theory is explained by a self-avoiding random vortex stretching mechanism. Among the consequences of this mechanism (see also [10]), the scale parameter of the log-stable turbulence PDF must follow:

$$\sigma_{\ln \varepsilon}^{\alpha} = \ln\left(\frac{\lambda}{\eta}\right), \; \alpha = 1.70 \tag{24}$$

Remembering the values of the Taylor and Kolmogorov length scales in section 4,

$$\sigma_{\ln \varepsilon} = 1.77 \tag{25}$$

This value of the scale parameter $\sigma_{\ln \varepsilon}$ is to compare to the result of the fitting of the size PDF in Figure 10. As indicated in the caption of Figure 10, the scale factor of the fitted log-stable law is given by:

$$\sigma_{\ln d} = 0.88 \tag{26}$$

From Eq.(25) and (26), it appears that:

$$\sigma_{\ln d} \approx \tfrac{1}{2} \sigma_{\ln \varepsilon} \tag{27}$$

It is actually quite difficult to put relation (27) to the test without modifying the atomization regime which is tightly linked to turbulence. Intriguingly enough this relation has also been obtained for another atomization mechanism without rotation [22]. In this work a high flow rate industrial nozzle is tested and the flow rate is high enough for the air to be entrapped in the flow so that no slowdown of droplets is observed. The smallest droplets are shown to follow a log-stable law of stability parameter 1.68 close to the value obtained for non-helical turbulent intermittencies. Large scales of the flow are governed by the classical Rayleigh-Taylor instability (in a non rotating framework). Eq.(27) implies the following relation between the droplet diameter and the local turbulent energy dissipation, seen as random variables:

$$d \propto \varepsilon^{\tfrac{1}{2}} \tag{28}$$

At this stage, it should be mentioned that this relation cannot be explained by the Hinze modeling which equates turbulent dynamic pressure to surface tension [23] and states that the maximal droplet size for an isolated droplet in a isotropic turbulent flow field is given by:

$$d_{max} = 0.725 \left( \frac{\gamma}{\rho_G} \right)^{3/5} \langle \varepsilon \rangle^{-2/5} \qquad (29)$$

In the present case, this expression would lead to $d_{max}$ equals to 500 $\mu m$ which is far from the maximal droplet size observed experimentally. Even worse, size and dissipation seems to be inversely correlated (note the minus sign in -2/5). For this reason, no refinement of this modeling seems to be able to explain the observed dependency (28).

As an attempt to explain Eq.(28), a possible scenario has been devised. Droplets are assumed to result from the reorganization of filaments. While this assumption was formulated for the first time a long time ago [18], the idea has been used again recently in [5,12,13]. In these works the agitation of the filaments introduces a reordering of the matter in a self-convolution process. The resulting droplet diameters are then considered as the random sums of the diameters of sets of elementary blobs that are developing at random locations on the filaments. The number of elementary blobs that merge to form a droplet is related to the corrugation of the filament. The width distribution of the filaments is assumed to be exponential using the maximum entropy formalism [13]. Since the convolution of several exponentials leads to gamma distributions, the author adequately fit their data with gamma PDF. However, the flow velocity in their experiments was possibly too low for turbulence to be a driving mechanism of atomization and the reported size PDF were quite narrow in comparison to those observed in this study.

The direct application of the previous model proved to be not feasible for the experimental results obtained in the present study. There was simply no way to find a gamma PDF that fits adequately the experimental in contrast to the log-stable law used in Figure 10.

While retaining the key aspects of the previous description (appearance of filaments and blobs leading to droplets), a different mechanism of agglomeration can be proposed to obtain eq.(28).

As previously, filaments that are developing from the liquid phase are expected to resist to the mixing with the surrounding air up to a certain point where they ultimately recess due to an agglomeration process of the blobs, at the end of which filaments form into droplets. A possible way to obtain the size distribution $n(d,t)$ of the droplets resulting from such an agglomeration process is given by Smoluchowski's equation [24]:

$$\frac{\partial n(d,t)}{\partial t} = \tfrac{1}{2} \int_0^d A(d-\xi,\xi) n(d-\xi,t) n(\xi,t) d\xi \\ - n(d,t) \int_0^\infty A(d,\xi) n(\xi,t) d\xi \qquad (30)$$

where $A$ is called the aggregation kernel. If $A$ is supposed constant this equation has an analytical solution [25] and the moments $m_i$ of order $i$ of the PDF, are given by:

$$m_0(t) = n(t) = \frac{2n_0}{2 + A n_0 t}, \qquad (31)$$

$$m_1(t) = d_0 (n_0 - m_0(t)), \qquad (32)$$

where $d_0$ is the size of the elementary blobs and $n_0$ is number of elementary blobs. The average droplet size is given by:

$$d = \frac{m_1(t)}{m_0(t)} = d_0 \frac{A n_0 t}{2} \qquad (33)$$

The overall aggregation time $t$ can be assumed to be the lifetime of large turbulent eddies i.e. their eddy turnover time or integral time scale

$$\tau_{int} = \frac{L_{int}}{k^{\frac{1}{2}}} \approx 42 \mu s \qquad (34)$$

whereas, the aggregation frequency $A$ can be thought to be inversely proportional to either the Taylor time scale

$$\tau_\lambda = \sqrt{15 \frac{v_G}{\varepsilon}} \approx 15 \mu s \qquad (35)$$

or to the Kolmogorov time scale

$$\tau_\eta = \sqrt{\frac{v_G}{\varepsilon}} \approx 4 \mu s \qquad (36)$$

Using, either (35) or (36), and (34) in (33), leads to the desired dependency (28):

$$d = \frac{m_1(t)}{m_0(t)} \approx d_0 \frac{n_0}{2} \frac{\tau_{int}}{\tau_\eta} = \frac{1}{2} \frac{L_{int}}{k^{\frac{1}{2}}} \sqrt{\frac{\varepsilon}{v_G}} \propto \varepsilon^{\frac{1}{2}} \qquad (37)$$

The main hypothesis made in the present model is that aggregation is a fast enough process to occur at a very high speed, $1/\tau_\eta$, during an overall time $\tau_{int}$. During this aggregation steps, the size of the resulting droplet increases as value of turbulence dissipation increases but decrease as turbulence intensity is increased. Other turbulent structures, possessing different values of the turbulence dissipation will lead to different size of droplets. The resulting droplet size will be proportional to the inverse of Kolmogorov time scale i.e. to the square root of the turbulence dissipation. In a way this can be considered as a turbulent micro mixing mechanism [26].

**Conclusion**

In this work an analysis of the atomization mechanism of a turbulent simplex pressure swirl full cone atomizer has been developed. The eccentric length of the feeding channel with respect to the vertical axis creates some angular momentum which is thereafter amplified by the contraction ratio of the funnel. Momentum conservation predicts a very high angular momentum which is confirmed by CFD computation. The latter bring evidences that a large boundary layer forms inside the nozzle. This slow boundary layer forms a conical sheet outside the nozzle. The behavior of the sheet is closer to a pressure swirl hollow cone atomizer and the breakup length of the sheet is adequately predicted by Dombrowski and Hooper empirical law. The inner vertical core atomize due to centrifugal force. It destabilizes according to linear instability theory. However comparison with prediction of Kubitschek and Weidman is not conclusive. This mechanism cannot predict the appearance of droplets smaller than one fifty micrometers. Since they are quite numerous another mechanism seems to be at work. By taking care to set aside small particles which have been strongly slowed down in the near nozzle area, the statistics of these small droplets have been shown to follow a log-stable PDF. The scale parameter of the Lévy stable law for the droplet magnitudes has been shown to be equal to half the value of the scale parameter of turbulence intermittency. Since this seems not to be unique, a mechanism based on the turbulent agglomeration and reorganization of elementary filaments has been devised in a tentative to explain this relationship. However what is still unexplained is the value of the stability index of the Lévy stable law. It is equal to 1.35 which is very different from the value 1.70 established for non helical homogeneous turbulence. Since the flow is strongly helical, it may suggest that another mechanism (different from the self-avoiding random vortex stretching mechanism) may be at work. Actually to cite [27]:

"All in all, we would expect turbulence in a rapidly rotating system to be very different from conventional turbulence"

**Acknowledgements**

The authors would like to thank Steen Gaardsted and Norskov Preben from Danfoss Burner Components, Nordborg, Denmark, for their help concerning the CAD/CFD part of this work and P$^r$. Emmanuel Plaut for his help concerning the linear instability computations.

1 Rimbert N. and Delconte A. *Solving Williams spray equation with a self-similar fragmentation kernel through classes method and quadrature method of moment* proceedings of the International Conference on Multiphase Flow.-2007 (Leipzig, Germany )

2 Kubitschek J.P. and Weidman P.D. *The effect of viscosity on the stability of a uniform rotating column in 0 gravity* J. Fluid. Mech. **572**, 261-286 (2007)

3 Kubitschek J.P. and Weidman P.D. *Helical instability of a rotating viscous liquid jet* Phys. Fluids **19**, 114108-1 (2007)

4 Gorokhovski M. and Herrmann M. *Modeling Primary Atomization* Ann. Rev. Fluid Mech., **40**: 343-366 (2008)

5 Eggers, J. and Villermaux, E., *Physics of liquid jets*, Reports on Progress in Physics 71, pp. 036601 (2008)

6 Rimbert N. and Séro-Guillaume O. *Log-Stable laws as asymptotic solutions to a fragmentation equation: application to the distribution of droplets in a high Weber number spray* Phys. Rev. E**. 69**, 056316 (2004)

7 Kida S. *Log-Stable distribution and Intermittency of Turbulence* J. Phys. Soc. of Jap. **60**, 1, 5-8, (1991),

8 Kida S. *Log-stable distribution in turbulence* Fluid Dyn. Res. **8,** 135-138, (1991)

9 Schertzer D., Lovejoy S., Schmitt F., Chigirinskaya Y. and. Marsan D. *Multifractal cascade dynamics and turbulent intermittency* Fractals, **5**, 3, 427-471 (1997)

10 Rimbert N. and Séro-Guillaume O. *Extension of Kida's log stable law in turbulence* C.R. Mecanique **331** (2003)

11 Rimbert N. *Simple model for turbulence intermittencies based on self-avoiding random vortex stretching*, Phys. Rev. E **81**, 056315 (2010)

12 Villermaux E., Marmottant Ph., and Duplat J. *Ligament-mediated spray formation* Phys. Rev. Lett., **92**, 7, 074501-1, (2004)

13 Marmottant Ph., *Atomization d'une liquide par un courant gazeux.* Thèse de l'université Joseph Fourier, 2001

14 S.A. Orszag, V. Yakhot, W.S. Flannery, F. Boysan, D. Choudhury, J. Maruzewski, and B.Patel. *Renormalization Group Modeling and Turbulence Simulations. In* International Conference on Near-Wall Turbulent Flows, Tempe, Arizona, 1993

15 Launder B.E., Reece G.J. et Rodi W. *Progress in the development of a Reynolds-stress turbulence closure* J. Fluid Mech. **68**, 3, 537-566 (1974)

16 Dombrowski N. Hooper P. *The Effect of Ambient Density on Drop Formation in sprays* Chem. Eng. Sci, **17**, 291, (1962)

17 Albrecht H.E., Borys M., Damaschke N., Tropea C. *Laser Doppler and Phase Doppler Measurement techniques*, Springer-Verlag (2003)

18 Lefebvre, H. *Atomization and Sprays* Hemisphere Publishing Corporation (1989)

19 Hsiang L.-P. and Faeth G.M. *Near-Limit Drop Deformation and Secondary Breakup* Int. J. Multiphase Flow, 18, 5, 635-652, (1992)

20  Tennekes H. and Lumley J.L. *A first Course in Turbulence* MIT Press (1972)

21  Hocking L.M. and Michael D.H. *The stability of a column of rotating liquid* Mathematika 6, **25** (1959)

22  Rimbert N. and Castanet G. *Evidences of turbulent cascading atomization in the bag-breakup regime* Proceedings of the ILASS – Europe 2010, 23rd Annual Conference on Liquid Atomization and Spray Systems, Brno, Czech Republic, September 2010

23  Hinze J.O. *Fundamentals of the Hydrodynamics Mechanism of Splitting in Dispersion Processes* AICHE J., 1, 289-295 (1955)

24  Smoluchowski, M. *Studies of Molecular Statistics of Emulsions and their Connection with Brownian Motion* Sitzungsberichte. Abt. 2a, Mathematik, Atronomie, Physik, Meteorologie und Mechanik ,123, 2381-2405, (1914)

25  Ramkrishna D. *Population Balances Theory and Applications to Particulate Systems in Engineering* Academic Press (2000)

26  Fox R.O. *Computational Models for Turbulent Reacting Flows* Cambridge University Press, (2003)

27  Davidson P.A. *Turbulence An introduction for scientists and engineers* Oxford (2004)